\documentstyle[12pt]{article}
\def\selecm{\tilde e^-}
\def\selecp{\tilde e^+}

\def\selecR{\tilde e_{\rm R}}
\def\selecL{\tilde e_{\rm L}}

\def\photino{\tilde\gamma}

\begin{document}
\pagestyle{plain}
\renewcommand{\thefootnote}{*)}

\title{\bf
\begin{quote}
 \raggedleft TMCP--95--1\\
 February 1995\\ ~
\end{quote}
The SUSY-GRACE system}
\font\fonts=cmbx12
\author{~\\
\bf Masato J{\fonts IMBO}\\
\it Computer Science Laboratory, Tokyo Management College,\\
\it Ichikawa, Chiba 272, Japan\\
\it (e-mail: jimbo@kekvax.kek.jp)\\
\and \bf Hidekazu T{\fonts ANAKA}\\
\it Faculty of General Education, Rikkyo University,\\
\it Tokyo 171, Japan\\
\it (e-mail: tanakah@kekvax.kek.jp)\\
\and \bf Toshiaki K{\fonts ANEKO}\\
\it Faculty of General Education, Meiji-gakuin University,\\
\it Totsuka, Yokohama 244, Japan\\
\it (e-mail: kaneko@minami.kek.jp)\\
\and \bf Tadashi K{\fonts ON}\\
\it Faculty of Engineering, Seikei University,\\
\it Musashino, Tokyo 180, Japan\\
\it (e-mail: kon@ge.seikei.ac.jp)\\
\and \bf M{\fonts INAMI}-T{\fonts ATEYA} collaboration\\
}\date{ }
\maketitle
\begin{abstract}
    We introduce a new method to treat Majorana fermions
on the GRACE system which has been developed for
the computation of the matrix elements for the processes
of the standard model.  In the standard model, we already
have such particles as Dirac  fermions,  gauge
bosons and scalar bosons in the system.   On the other
hand, in the SUSY models there are Majorana fermions.
    In the first instance, we have constructed a system
for the automatic computation of cross-sections for the
processes of the SUSY QED.   It is remarkable that our
system is also applicable to another model including
Majorana fermions ({\it e.g.} MSSM) once the definition of
the model file is given.
\end{abstract}
\renewcommand{\thefootnote}{\sharp\arabic{footnote}}
\renewcommand{\theequation}{\arabic{section}.\arabic{equation}}
\setcounter{footnote}{0}

\section{Introduction}

    It has been widely accepted that there exists a symmetry called
supersymmetry (SUSY) between bosons and fermions at the unification-energy
scale.  SUSY, however, is broken at the electroweak-energy scale.
The relic of SUSY is expected to remain as a rich spectrum of SUSY particles,
partners of usual matter fermions, gauge bosons and Higgs scalars, named
sfermions, gauginos and higgsinos, respectively~\cite{theor-a,theor-b,theor-c}.

    Since there exist so many particles and their interactions, it is a
skilled job to calculate the cross-sections for the processes with the final
3-body or more.  We have already known within the standard model that the
numerical calculation of the helicity amplitudes with the program package
CHANEL~\cite{chan} is more advantageous to such a case than that of the traces
for the gamma matrices with REDUCE.

    It, however, is also hard work to construct a program with many
subroutine calls of CHANEL.  Thus we need a more convenient way to carry out
such a work.  The GRACE system~\cite{pre-grc}, which automatically generates
the source code for CHANEL, is one of the solutions.  The system also includes
the interface and the library of CHANEL, and the multi-dimensional integration
package BASES~\cite{bas}.

    The GRACE system has been developed for the computation of the matrix
elements for the processes of the standard model.  In the standard model,
we already have such particles as Dirac fermions, gauge bosons and scalar
bosons in the system.   On the other hand, in the SUSY models there are
Majorana fermions as mixed states of neutral gauginos and higgsinos.

    Thus we are able to computate the SUSY processes with an automatic system,
provided that we develop an algorithm to treat Majorana fermions in the system.
The aim of this work is to construct such a system and test it with REDUCE.
 Since anti-particles of Majorana fermions are themselves, there exists
so-called `Majorana-flip', the transition between particle and anti-particle.
This is the most important problem which we should solve when we realize the
automatic system for computation of the SUSY processes.

\section{SUSY-CHANEL into new GRACE}

   The method of computation in the program package CHANEL is as follows:
\begin{enumerate}
  \item To divide a helicity amplitude into vertex amplitudes.
  \item To calculate each vertex amplitude numerically as a complex number.
  \item To reconstruct of them with the polarization sum, and calculate
  the helicity amplitudes numerically.
\end{enumerate}
The merit of this method is that the extension of the package is easy,
and that each vertex can be defined only by the type of concerned particles.

    Here we propose an algorithm for the implementation of the embedding
Majorana fermions in CHANEL as follows:
\begin{itemize}
  \item \underline{\bf policy}
  \begin{enumerate}
    \item To calculate a helicity amplitude numerically, and square the
    absolute value of it.
    \item To replace each propagator by wave functions, and calculate
    vertex amplitudes.
    \item \underline{\bf Not to} move charge-conjugation matrices.
  \end{enumerate}
  \item \underline{\bf method}
  \begin{enumerate}
    \item To choose a direction on a fermion line.
    \item To put wave functions, vertices and propagators along the direction
    in such a way:
    \begin{itemize}
      \item[~i)] To take the transpose for the reverse direction of
      fermions
      \item[ii)] To use the propagator with the charge-conjugation
      matrix for\\
      the Majorana-flipped one.
    \end{itemize}
  \end{enumerate}
\end{itemize}
As a result, the kinds of the Dirac-Majorana-scalar vertices are limited to
four types:
\begin{itemize}
\begin{itemize}
  \item[(1)] $\overline{U} \Gamma U$
  \item[(2)] $U^{\rm T} \Gamma \overline{U}~^{\rm T}$
  \item[(3)] $\overline{U} C^{\rm T} \Gamma^{\rm T} \overline{U}~^{\rm T}$
  \item[(4)] $U^{\rm T} \Gamma^{\rm T} C^{-1} U$
\end{itemize}
\end{itemize}
where $U$'s denote wave functions symbolically without their indices, and $C$
is the charge-conjugation matrix. The symbol $\Gamma$ stands for the scalar
vertex such as
\[ \Gamma = A_{\rm L}\cdot{{1 - \gamma}\over{2}} +
A_{\rm R}\cdot{{1 + \gamma}\over{2}} ~~. \]

    The vertices (2)$\sim$(3) are related to the vertex (1) which is the same
as the Dirac-Dirac-scalar vertex in the subroutine of CHANEL.  Thus we can
build three new subroutines for the added vertices.

    On the other hand, the GRACE system becomes more flexible for the extension
in the new version called `{\bf grc}'~\cite{grc-pp}.  We have performed the
installation of the subroutines above with the interface on the new GRACE
system.

\section{Numerical results for tests}

    At the start for the check of our system, we have written the model file
of SUSY QED.  In this case, there is only one Majorana fermion, photino.
The tests have been done in such a manner:
\begin{itemize}
  \begin{enumerate}
    \item To calculate the differential cross-sections at a point of the
    phase space in the two methods with GRACE and REDUCE.
    \item To calculate the differential cross-sections over the
    phase space with REDUCE.
    \item To integrate the differential cross-sections over the
    phase space in the two methods with GRACE and REDUCE.
  \end{enumerate}
\end{itemize}
For the GRACE system, we can get the differential cross-sections and the
scattered plots by one time of the integration step with BASES.

    In Table I, the tested processes are shown as a list.
It has been also confirmed that the cross-sections for the charge-conjugated
processes $e^- e^- \rightarrow \selecm \selecm$ and $e^+ e^+ \rightarrow
\selecp \selecp$ are identical.  The references in the table are not the
results of the tests, but for help.  Hereafter the typical results are
presented at the JLC-I energy.  The masses are $M_{\photino} = 50$ GeV,
$M_{\selecR} = 60$ GeV and $M_{\selecL} = 70$ GeV.

\begin{table}
\begin{center}
  \begin{tabular}{llclc}  \hline
  Process &  &  Number of diagrams  &  Comment  &  Reference  \\  \hline\hline
$e^- e^- \rightarrow$ & $\selecR^- \selecR^-$ & 2 & Majorana-flip & \\
                    & $\selecL^- \selecL^-$ & 2 & in internal lines & \\
                    & $\selecR^- \selecL^-$ & 2 & & \\  \hline
$e^- e^+ \rightarrow$ & $\selecR^- \selecR^+$ & 2 & Including pair &
\cite{mj}\\
               & $\selecL^- \selecL^+$ & 2 & annihilation & \cite{mj} \\ \hline
$e^- e^+ \rightarrow$ & $\selecR^- \selecL^+$ & 1 & Values are & \cite{mj} \\
                    & $\selecR^+ \selecL^-$ & 1 & equal & \cite{mj} \\  \hline
$e^- e^+ \rightarrow$ & $\photino \photino$ & 4 & F-B symmetric & \\  \hline
$e^- e^+ \rightarrow$ & $\photino \photino \gamma~$ & 12 & Final 3-body
   & \cite{tk} \\ \hline
  \end{tabular}
\end{center}
  \hspace*{4cm} Table~I. The list of the tested processes.
\end{table}%

    In Fig.~1, we show the angular distribution of the process
$e^{\mp} e^{\mp} \rightarrow \selecL^{\mp} \selecL^{\mp}$.  In this case,
there exist two Feynman diagrams with Majorana-flip in the internal lines.
Results from two methods exactly coincide.

%

    In Fig.~2, we show the angular distribution of the process
$e^- e^+ \rightarrow \photino \photino$.  In this case, there exist four
Feynman diagrams with Majorana-flip in the external lines, and with both
selectrons ($\selecR$ and $\selecL$) in the internal lines.  Here we use BASES
for the calculation from the REDUCE output. The result is in beautiful
agreement with the value that is obtained by GRACE at each bin of the
histogram.

%

\section{Summary}

    We introduce a new method to treat Majorana fermions on the GRACE system
for the automatic computation of the matrix elements for the processes of the
SUSY models.  In the first instance, we have constructed the system for the
processes of the SUSY QED because we should test our algorithm with the
simplest case.  The numerical results convince us that our algorithm is
correct.

    It is remarkable that our system is also applicable to another model
including Majorana fermions ({\it e.g.} MSSM) once the definition of the model
file is given.  We should computate the single-photon event~\cite{tk} and
the resultant single-electron (positron) event from the single-selectron
production~\cite{spu} as soon as possible.  It should be emphasized that the
GRACE system including SUSY particles is the powerful tool for the purpose.

\section{Acknowledgements}

  This work was supported in part by the Ministry of Education,
  Science and Culture, Japan under Grant-in-Aid for International
  Scientific Research Program No.04044158. Two of us (H.T. and M.J.)
  have been also indebted to the above-mentioned Ministry under Grant-in-Aid
  No.06640411.


\begin{thebibliography}{99}
  \bibitem{theor-a} H.P. Nilles, {\sl Phys. Rep.} {\bf 110} (1984), 1.\\
   H.E. Haber and G.L. Kane, {\sl Phys. Rep.} {\bf 117} (1985), 75.\\
   M. Chen, C. Dionisi, M. Martinez and X. Tata, {\sl Phys. Rep.}
  {\bf 159} (1988), 201.
  \bibitem{theor-b} R. Barbieri, {\sl Riv. Nuovo Cimento} {\bf 11} (1988).
  \bibitem{theor-c} R. Barbieri {\it et al.}, {\sl Z PHYSICS AT LEP 1},
  CERN Report CERN 89-08 (1989) Vol. 2, p.121.
  \bibitem{chan} H. Tanaka, {\sl Comput. Phys. Commun.} {\bf 58} (1990), 153.
  \bibitem{pre-grc} T. Kaneko, in {\sl New Computing Techniques in Physics
  Research}, edited by D. Perret-Gallix and W. Wojcik, {\' E}dition du CNRS,
  Paris, 1990, p.555.\\
   T. Kaneko and H. Tanaka, in {\sl Proceedings of the Second Workshop
  on Japan Linear Collider (JLC)}, KEK, November 6-8, 1990, edited by
  S. Kawabata, KEK Proceedings 91-10 (1991), p.250.\\
   T. Kaneko, in {\sl New Computing Techniques in Physics Research II},
  edited by D. Perret-Gallix, World Scientific, Singapore, 1992, p.659.\\
   T. Ishikawa {\it et al.}, Minami-Tateya group, in {\sl GRACE manual},
  KEK Report 92-19, 1993.\\
   and References therein.
  \bibitem{bas} S. Kawabata, {\sl Comput. Phys. Commun.} {\bf 41} (1986), 127.
  \bibitem{grc-pp} T. Kaneko, preprint KEK-CP-020 (KEK Preprint 94-83 /
  MGU-CS/94-01), (1994).
  \bibitem{mj} M. Jimbo, in {\sl Proceedings of the Second Workshop on Japan
  Linear Collider (JLC)}, KEK, November 6-8, 1990, edited by S. Kawabata, KEK
  Proceedings 91-10 (1991), p.185.\\
   M. Jimbo, {\sl Memoirs of Tokyo Management College}, {\bf I} (1993), 101.\\
   and References therein.
  \bibitem{tk} T. Kon, {\sl Prog. Theor. Phys.}, {\bf 79} (1988), 1006.\\
   and References therein.
  \bibitem{spu} M. Jimbo, T. Kon and T.Ochiai, Rikkyo University preprint
  RUP-87-1 (1987).\\
   M. Jimbo, {\sl Prog. Theor. Phys.}, {\bf 79} (1988), 899.\\
   and References therein.
\end{thebibliography}
\end{document}